\documentclass{aa}
\usepackage{graphicx}

\begin{document}  

\thesaurus{ ; ; ; ; ; }

\title{High resolution images at 11 and 20 $\mu$m of the Active 
Galactic Nucleus in NGC\,1068  
\thanks{ Based on observations collected at the ESO/La Silla 3.6m telescope, 
Proposal 62.P-0445, and at the CFHT/Hawaii 3.6m telescope.}} 
\author{ D.\,Alloin \inst{1} \and E.\,Pantin \inst{2} \and P.O.\,Lagage 
\inst{2} 
\and G.L.\,Granato \inst{3} }
\institute{ European Southern Observatory, Casilla 19001, 
Santiago, Chile
\and  DSM/DAPNIA/Service d'Astrophysique, CEA/Saclay, F-91191
 Gif-sur-Yvette, France
\and  Osservatorio Astronomico di Padova, I-35122 Padova, Italy}
\offprints{D. Alloin} 
\mail{dalloin@eso.org}
\date{Submitted to Astronomy \& Astrophysics}
\titlerunning{NGC\,1068 high resolution imaging at 11 and 20 $\mu$m}
\maketitle

\begin{abstract}

We present  diffraction-limited IR images at 11.2 and 20.5 $\mu$m of the  
central $6'' \times 6''$ region in NGC\,1068, collected with the 
CAMIRAS
instrument mounted at the f/36 IR focus of the CFHT/Hawaii 3.6m
telescope and at the f/35 IR focus of the ESO/La Silla 3.6m 
telescope, respectively. After 
deconvolution, the achieved resolution ($0.6''$) reveals a prominent 
central
core emitting about 95\% of the total flux at these wavelengths, as 
well as extended 
emission, to the South-West (PA = 210\degr) and to the North-East 
(PA = 35\degr), broken into patchy components which are particularly 
conspicuous at 20.5 $\mu$m and can be isolated as
 individual clouds. The central core shows an East-West FWHM 
 of $0.6''$ (hence unresolved) and a North-South FWHM of $0.9''$
 corresponding to a resolved full size extension of 
 $\approx$ 100 pc. Such an 
 elongated 
 shape is in agreement with model predictions of a dusty/molecular torus
 surrounding the central engine in NGC\,1068, observed under an inclination 
 angle
 of around 65\degr.
Considering that the core at 11.2 and 20.5 $\mu$m is coincident 
with the 
core also seen at 2.2, 3.5 and 4.8 $\mu$m and that this features the
location of the central engine (also the radio source S1), the 
extended mid-IR 
emission is found to follow closely the radio jet--like structure in 
both the North-East and South-West quadrants. In the North-East quadrant, 
we observe that the mid-IR emission arises predominantly on the eastern 
side 
of the ionizing cone defined by the HST NLR [OIII]--emitting clouds and 
is still quite prominent at the wide base of the so-called Northeast radio 
lobe,
$4''$ away from the central engine. A detailed comparison of the 
extended mid-IR emission with model
predictions requires that future AGN modeling includes both a molecular/dusty
 torus and a distribution of material away from the equatorial plane of
 the torus, i.e. in and around the NLR. 
 

\end{abstract}

\section{Introduction}

The current unification scheme for Active Galactic Nuclei (herefater AGN)
postulates the existence of a thick molecular/dusty torus surrounding a
black-hole/accretion-disc system (see, e.g. Krolik 1999). Attempts to 
unveil this
torus in a nearby AGN like NGC\,1068 have drawn considerable effort on the 
side 
of high spatial resolution imaging, in particular 
in the near-IR (Chelli et al. 1987,  
Gallais, 1991, Young et al. 1996, Marco et al. 1997, Thatte et al. 
1998, Rouan et al. 1998, Alloin et al. 1998, Wittkowski et al. 1998, 
Weinberger, Neugebauer \& Matthews 1999, Marco \& Alloin, 2000) and 
mid-IR (Braatz et
al. 1993, Cameron et al. 1993, Bock et al. 1998, Lumsden et al 1999). 
With respect to the presence of such a torus, the most compelling results 
are: (a) VLBA 8 GHz data (Gallimore et al. 
1997) revealing a 1 pc elongated distribution of 
ionized gas  at PA $\approx$ 110\degr\ (the inner "hot zone" of the torus?),
 (b) adaptive optics (AO) 
observations at 2.2 $\mu$m (Rouan et al, 1998) showing extended emission 
along PA $=$ 102\degr\ and up to a 15 pc radius (in addition to a prominent 
unresolved core with size less that 9 pc and in addition to another 
20 pc radius extended 
emission along PA $=$ 15\degr\ which follows the Narrow Line Region) and
at 3.5 and 4.8 $\mu$m (Marco \& Alloin, 2000) revealing structures similar 
in shape/orientation and on comparable scales, (c) near- and mid-IR imaging
polarimetry (Lumsden et al. 1999) calling for the presence of hot and
warm dust components fully consistent with those detected with AO. The 
structure elongated at PA $\approx$ 102\degr\ 
could trace the equatorial plane of an inclined dusty/molecular torus, 
from its
inner walls of ionized material seen in radio, to its outer
parts of thermally emitting dust with a temperature range from 
T $\approx$ 1500 K (dust sublimation temperature) to a few 100 K.

Models by Krolik \& Begelman (1986), Pier \& Krolik (1993), Efstathiou  
\& Rowan-Robinson (1994), Granato \& Danese (1994), 
and Granato, Danese \& Franceschini  (1997)
have explored torus sizes from 1 to 100 pc. They predict emission in 
the near-IR and mid-IR and provide high 
resolution model-maps at various wavelengths. Therefore, it
is particularly timely to obtain high resolution mid-IR 
images to compare with model-maps. 

Here, we present and discuss new results from diffraction-limited
images obtained with the CFHT/Hawaii and ESO/La Silla 3.6m
telescopes, at 11.2 and 20.5 $\mu$m respectively. Hereafter, the distance 
used for NGC1068 is 14.4 Mpc, leading to the equivalence 
  1\arcsec $\approx$ 70 pc in the object. 

\section{ Data collection and Reduction}

The observations were performed using the CEA/SAp mid-IR camera CAMIRAS 
(Lagage et al 1993),
equipped with a Boeing 128x128 pixels BIB detector 
which is sensitive up to 28 $\mu$m.
The AGN in NGC1068 was observed with CAMIRAS attached to the 
ESO/La Silla 3.6m telescope on 1998 November 8-10 and to the CFHT/Hawaii 
3.6m telescope on 1999 August 1. During these two runs seeing and
weather conditions -- humidity and amount of atmospheric precipitable 
water -- were 
particularly favorable to observing in the mid-IR and extremely stable.

The orientation of the array on the
sky was carefully determined at the start of each observing run.
The pixel size was $0.315''$ (20 $\mu$m window, ESO/La Silla) and 
$0.29''$ (11 $\mu$m window, CFHT/Hawaii).

In the 20 $\mu$m window, we used a filter centered at 20.5 $\mu$m and with a 
bandpass (FWHM) $\Delta\lambda$ =
1.11 $\mu$m. This filter is free of any important line contribution.
In the 11 $\mu$m window, we used a filter centered at 11.2 $\mu$m and
with a bandpass $\Delta\lambda$ = 0.44 $\mu$m: although the 11.3 $\mu$m 
PaH line emission falls within this filter, one should notice that the
spectroscopic results from the ISO satellite do not show its presence 
in the AGN of NGC\,1068 (aperture of $3''$). Therefore we are confident that 
these two filters reflect mostly the mid-IR continuum emission of the
AGN.
  To avoid saturation of the detector by the high ambient photon background, 
the image elementary integration times were chosen to be 9.1 msec 
and 19.1 msec, respectively at 20.5 and 11.2 $\mu$m. Standard chopping 
and nodding 
techniques were applied with a chopping throw of $20''$. Elementary images 
were coadded in real time during 3 sec chopping cycles. 
At 20.5 $\mu$m a total integration time of 22.5 min was spent on the source 
observed at an airmass less than 1.3.
At 11.2 $\mu$m, the total integration on source was of 65 min and
the mean airmass of 1.04. 
A shift-and-add procedure was applied to the final images through each filter. 
The final signal-to-noise ratio at the peak emission on the images is 265 at 
20.5 $\mu$m and 980 at 11.2 $\mu$m.

\begin{table*}[t]
\caption{Position, size and flux measurements of the various structures 
identified at 20.5 $\mu$m, as derived from the deconvolved image (down to 
a final resolution of $0.6''$). The extension provided for the core 
corresponds to the FWHMs of its E-W and N-S profiles, while the extension
quoted for the clouds (a), (b), (c) and (d) corresponds to a full size
as measured from the map
at 20.5 $\mu$m. Fluxes given in this table have been obtained through 
a mask fitting the full extent of the core and of each of the clouds (see text).
The total flux found at 11.2 $\mu$m in 1999 is about 30\% larger than 
the figure 
quoted by Lumsden et al for the year 1995. According to Glass (1997) 
and Marco \& Alloin (2000), the 3.5 $\mu$m flux has been increasing 
steadily since 1974. Therefore, the 11.2 $\mu$m flux difference found 
between 1995 and 1999 is most probably related to an intrinsic flux increase
of the AGN.}
\begin{center}
\begin{tabular}{lcccc}
\hline

Component        &Extension      &Distance    &Flux 20.5$\mu$m         &Flux 
11.2$\mu$m  \cr
              &($''\times''$)    &($''$)    &(1$\sigma$ error, (Jy))    
&(1$\sigma$ error,(Jy)) \cr

Core          &$0.6\times 0.9$      &0          &57.5 (6)                 &37.5 
(2)       \cr

(a) N-E       &$1.8\times 4.0$      &3.7        &1.4 (0.1)                &0.15 
(0.01)   \cr

(b) N-E       &$1.3\times 1.3$      &1.8        &1.1 (0.1)                &2.6 
(0.15)    \cr

(c) S-W       &$1.3\times 1.6$      &2.1        &1.0 (0.1)                &0.3 
(0.02)   \cr     

(d) S-W       &$1.3\times 1.3$      &3.5        &0.15(0.01)               
&undetected    \cr
\hline
\end{tabular}
\end{center}
\label{parameters}
\end{table*}

The nearby reference star $\tau$ 4 Eri was frequently monitored to serve 
as a PSF and flux standard at 20.5 $\mu$m, while $\alpha$ Ceti was used at
11.2 $\mu$m (van der Bliek et al 
1996). The precision achieved in flux measurements is +/-- 5\%
and +/-- 10\%, respectively at 11.2 $\mu$m and 20.5 $\mu$m.

Because final images are limited by the seeing 
and above all by the 3.6m telescope diffraction-pattern 
(FWHM of $1.4''$ and $0.8''$ at 20.5 $\mu$m and 11.2 $\mu$m respectively),
applying a deconvolution procedure is mandatory. We use the 
Entropy Method developed by Pantin \& Starck (1996), based upon wavelet 
decomposition. The Shannon theorem (1948) prescription allows to persue
the deconvolution down to twice the pixel size,
in this case down to $0.6''$, and iterations are stopped according to
the residual map, whose properties should be consistent with data 
noise characteristics.

\section{Results and Discussion}

\subsection{Structure and flux contributions}
 
We provide in Figure 1 a set of the images at 20.5 $\mu$m: the NGC1068 image 
before deconvolution, the PSF stellar image,  and the NGC1068 image after 
deconvolution. 
A similar set of images at 11.2 $\mu$m is displayed in Figure 2. 
The general orientation of the extended emission runs through the 
North-East to South-West quadrants, with a noticeable North-South 
elongation of the inner isophotes. The deconvolved image and contours at
11.2 $\mu$m are fully consistent with the map at 12.4 $\mu$m obtained at
a comparable spatial resolution by Braatz et al (1993). In the 10 $\mu$m 
window, the speckle data (over a $5'' \times 5''$ 
region, with a resolution down to $0.2''$, but with a smaller dynamical
range) analyzed by Bock et al (1998) indicate however a different 
orientation with the extended emission in the inner $5'' \times 5''$ 
region running  
through the North-West to South-East quadrants: this does not appear to be
consistent with either Braatz et al (1993) or the new data set presented in 
this paper. This discrepancy remains to be elucidated.

\begin{figure*}[ht]
\resizebox{16cm}{!}{\includegraphics[scale=1.,bb=0cm 0cm 18cm 18cm,clip=true]{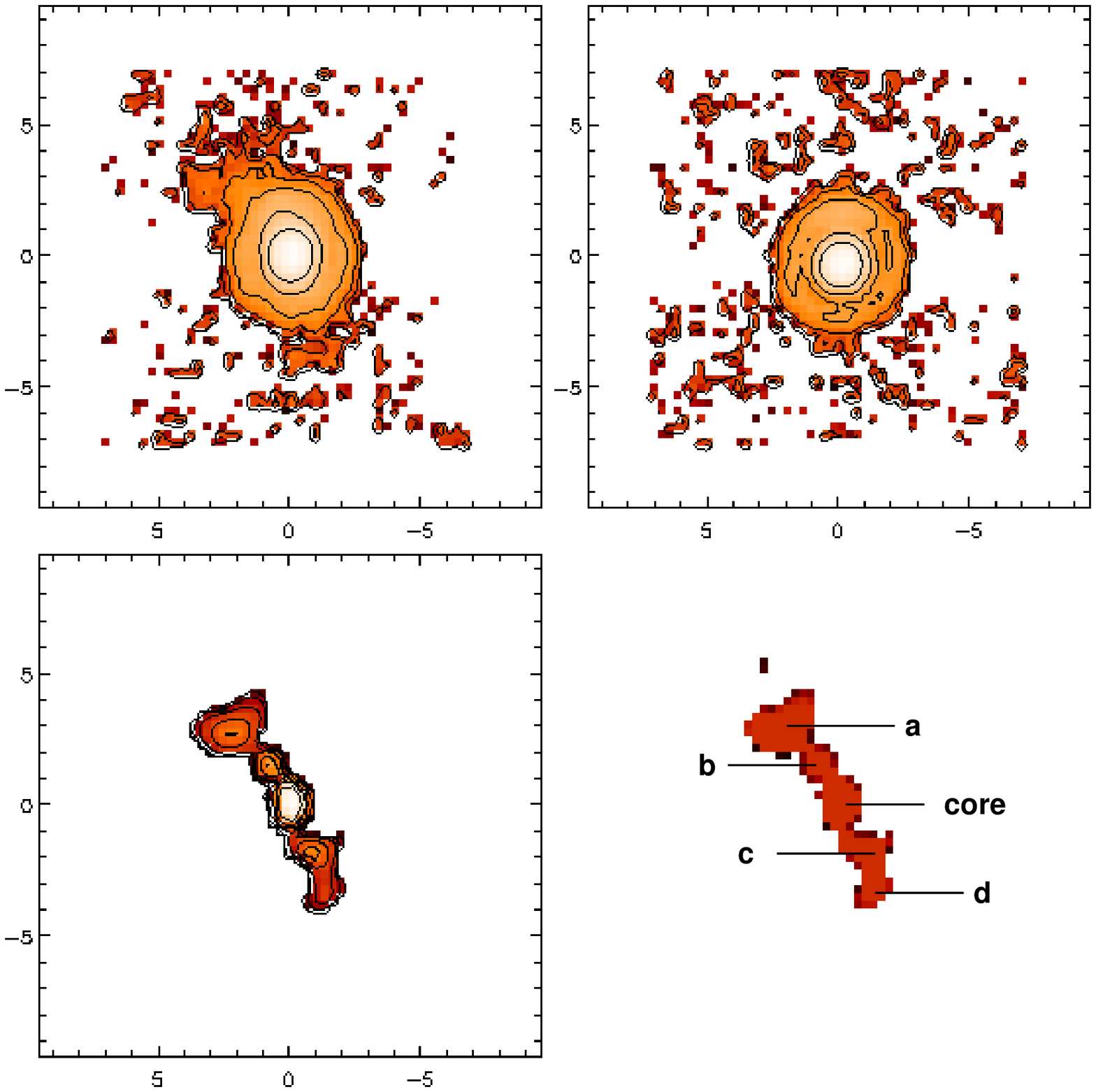}}
\caption{The 20.5 $\mu$m data. Top left : the NGC 1068 raw image. Top right : 
the
reference star used as PSF. Bottom left : the deconvolved 
image of NGC 1068. The bottom right image
shows a sketch of the 5 structures discussed
in the text, clouds (a),(b),(c),(d) and the core. On all images, North is 
to the top, East to the left. 
The pixel scale is $0.315''$/pixel, and the total field spans about 
$19.2 ''\times 19.2 ''$. On the raw image and the PSF image, a series 
of 6 contours (step by factor 3 in intensity from one to the next) 
has been superimposed. On the deconvolved image, 8 contours (same step)
have been superimposed.}
\label{fig1}
\end{figure*}

\begin{figure*}
\resizebox{16cm}{!}{\includegraphics[scale=1.,bb=0cm 0cm 18cm 18cm,clip=true]{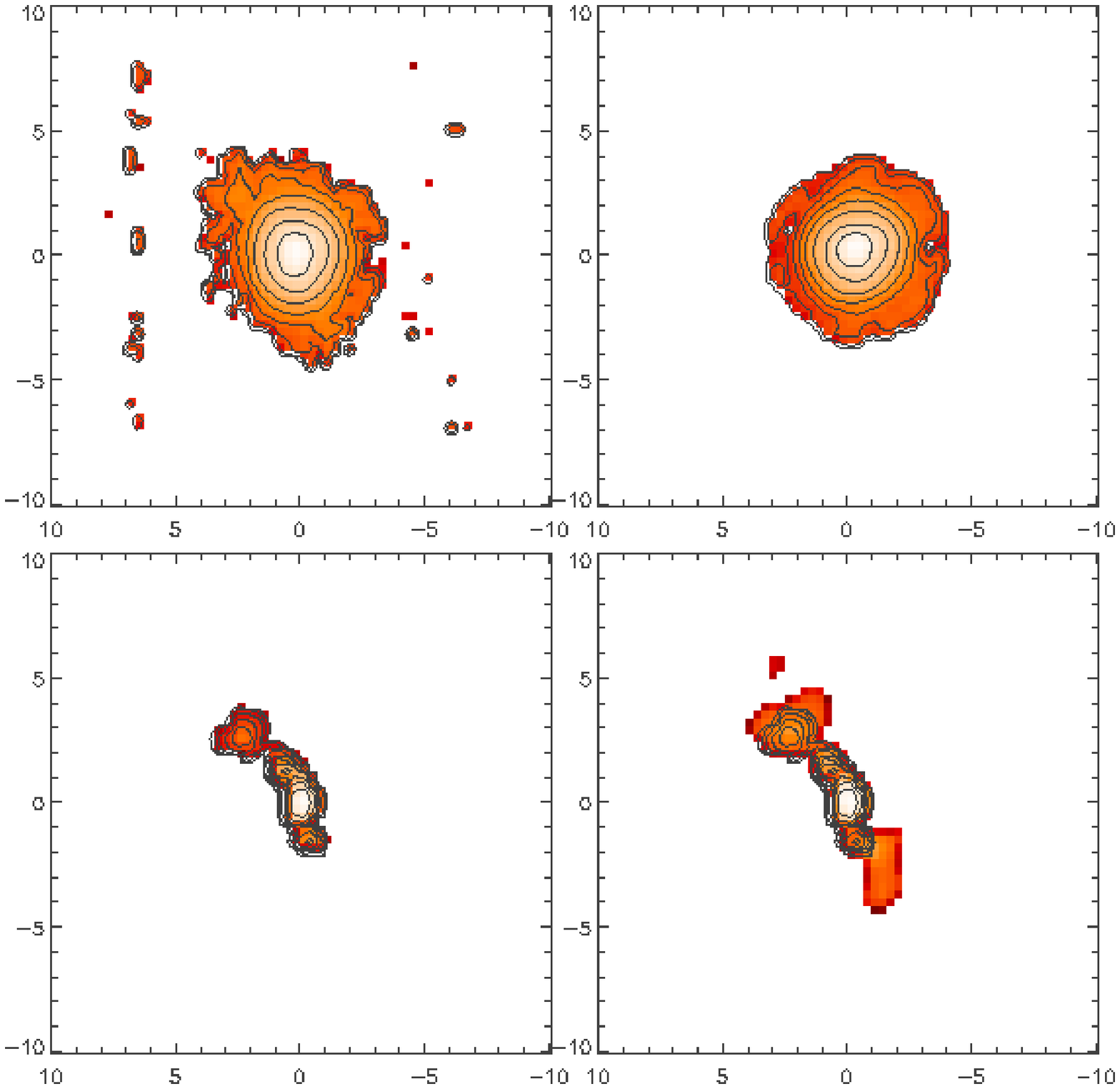}}
\caption{The 11.2 $\mu$m data. Top left : the NGC 1068 raw image. Top right : 
the
reference star used as PSF. Bottom left : the deconvolved 
image of NGC 1068. On all images, North is to the top, East to the left. 
The pixel scale is $0.29''$/pixel, and the total field spans  
$20''\times 20 ''$. On the raw image and the PSF image, a series of 8 
contours (step by factor 3 in intensity from one to the next) has 
been superimposed. On the deconvolved image, 10 contours (same step)
have been superimposed.
The bottom right panel features in its background
the 20.5 $\mu$m deconvolved image on top of which the 11.2 $\mu$m 
contours have been superimposed,
showing the correspondence of structures at 11.2 and 20.5 $\mu$m. Note 
that cloud (d) seen on the 20.5 $\mu$m image is 
not detected at 11.2 $\mu$m, probably because of temperature and opacity 
effects.}
\label{fig2}
\end{figure*}

The deconvolved images, both at 11.2 and 20.5 $\mu$m show the 
presence of a prominent central core. Cuts 
through the core along the North-South and East-West 
directions provide FWHM values of $0.9''$ and $0.6''$ respectively.
Therefore, the core remains unresolved in the East-West direction, 
while it appears to be extended North-South with intrinsic FWHM size 
of around 50 pc. Again this result is
in very good agreement with Braatz et al (1993).

Regarding the extended emission, one notices that conspicuous features 
can be seen further away from the core at 20.5 $\mu$m than at 11.2 $\mu$m.  
In the North-East quadrant we have identified two clouds, (a) and (b),
along PA $=$ 35\degr, at mean distances from the core of $3.7''$ and $1.8''$, 
respectively. 
Cloud\,(a), which is bright both at 11.2 and 20.5 $\mu$m, appears to 
widen perpendicularly to the main direction of the extended emission
and runs from PA $=$ 15\degr\ to PA $=$ 55\degr. In the South-West 
quadrant, the emission 
closest to the core extends along 
PA $=$ 210\degr\ and then aligns North-South: two clouds, (c) and (d), can
be singled out at mean distances from the core 
of $2''$ and $3.5''$. The results are summarized in Table 1.

Flux measurements at 11.2 and 20.5 $\mu$m have been performed using a mask
built from the 20.5 $\mu$m image and delineating the extent of the four 
clouds 
and of the core, so that the 11.2 $\mu$m to 20.5 $\mu$m flux 
ratio can be derived in a consistent fashion. The mask at the position of 
the core takes into account the full 
extent ($2''\times 1.2''$) of the core (rather than its FWHMs values
given in Table 1). The resultant fluxes 
are given in Table 1: one notices that the core contributes a 
very large fraction of the mid-IR emission, 95.6\% at 11.2 $\mu$m and 94\% at 
20.5 $\mu$m.

\subsection{Comparison with maps at other wavelengths}

We have just seen that the mid-IR emission extends away from the core 
up to a radius of about 
300 pc, both in the North-East and South-West quadrants. In order to compare 
this extension/structure to those observed at other wavelengths, in the radio 
range with the VLA or in 
the UV/optical with HST, it is mandatory to register
precisely the mid-IR map with respect to the near-IR, visible and radio ones. 

As we did not have the possibility to perform astrometric measurements 
through simultaneous 
observations in the mid-IR and near-IR or visible (as performed in 
Marco et al 1997),
 we are assuming firstly that the cores at 11.2 and 20.5 $\mu$m 
are coincident 
and secondly that their location also coincides with that of the cores 
observed at 4.8, 3.5 and 2.2 $\mu$m. Regarding the first assumption, one can
check from the model of NGC\,1068 developed 
by Granato et al (1997) which predicts the offsets between 
emission peaks at various wavelengths in the near-IR to mid-IR, that these
offsets remain in the range of a few hundredths of an arc second and
can be ignored. Regarding the second assumption, the cores observed at 
4.8, 3.5 and 2.2 $\mu$m have 
been shown to be positionally coincident indeed (Rouan et al, 1998, Marco \& 
Alloin 2000). 
And finally, Marco et al (1997) have obtained the astrometric positioning of 
the core at
2.2 $\mu$m, with respect to the HST map in the visible of the central area 
in
NGC\,1068: it was found to be located $0.28''$ South and $0.08''$ West of 
the so-called optical continuum peak seen with HST (Lynds et al 1991). The
study by Marco et al (1997) also allowed a precise registration of the 
2.2 $\mu$m map
with respect to the 12.4 $\mu$m map from Braatz et al (1993) and to available 
radio, optical and UV maps. The conclusion is that the core at 4.8, 3.5 and 2.2 
$\mu$m appears to be coincident with
the 12.4 $\mu$m peak position found by Braatz et al, the radio source S1, 
the symmetry center of polarization both in the UV/optical (Capetti 
et al 1995) and in the mid-IR (Lumsden et al 1999). We consider this
core to feature the central engine (hidden in UV/optical maps) 
and from all the above quoted studies we can ascertain the consistency 
of the assumptions made earlier.

We have compared the mid-IR maps at 11.2 and 20.5 $\mu$m to the radio maps and 
to the HST [OIII]--emitting cloud map, provided on a suitable scale in
Bland-Hawthorn et al (1997) and with a registration as above. We have 
reached the following conclusions:\\
-- compared to the radio maps from Gallimore et al (1996a,b) and Muxlow et al
(1996), the observed North-South elongation of the mid-IR core is reminiscent 
of the North-South alignment of the radio sources S2, S1 and C (see 
Figure 2 in Bland--Hawthorn et al 1997 for a synthesis of the radio 
source nomenclature after Wilson \& Ulvestad 1987). At the 
location of radio source C, the orientation of the radio jet-like
structure changes abruptly and becomes close to PA $=$ 35\degr, matching 
then the 
general direction 
of the mid-IR extended emission in the North-East quadrant. The mid-IR 
cloud\,(a) coincides perfectly with the wide base of the so-called
Northeast radio lobe. In the South-West
quadrant, the radio jet-like  
structure is along PA $=$ 210\degr\, again very well aligned with the mid-IR
extended emission. The 
so-called Southwest radio hotspot is situated about $4''$ away from the  
central 
engine, S1, and can be put in correspondence 
with the mid-IR cloud\,(d).\\
-- compared with the HST [OIII]--emitting cloud map, the
North-South elongation of the mid-IR core is found to overlap
to the North with 
HST cloud-A and cloud-B, two NLR clouds distributed along the North-South 
direction. The extended mid-IR emission in the North-East quadrant is 
along PA $=$ 35\degr\ and arises to the eastern edge of the ionizing
cone (which, on a large scale extends between PA $=$ --15\degr\ and 
PA $=$ 35\degr) defined by the 
[OIII]--emitting complexes, beyond cloud-F and to the East of cloud-G which
is itself located about $2''$ away from the central engine. It should be
noted as well that the mid-IR cloud\,(a) is located very close to
an optical emission knot early identified in the literature as the 
"Northeast knot"
in the direction PA $=$ 35\degr\ and at a distance about $4''$ from 
the central engine (Elvius 1978 for the first identification). The extended 
mid-IR emission in the South-West quadrant has no counterpart in the 
HST [OIII]--emitting cloud map, as expected from the heavy obscuration
on that side of the AGN. Therefore, it is
conspicuous that the extended mid-IR emission in the North-East quadrant
arises aside the NLR (as featured by the [OIII]--emitting clouds) and from 
material which is not 
directly exposed to the central ionizing source. However, the tight 
correlation existing between the mid-IR extended emission and the 4.9 GHz 
emission is
a result which signals the importance of the radio jet-like emission 
impacting on interstellar
material located above and below the equatorial plane of the 
dusty/molecular torus.

\subsection{Nature of the mid-IR core}

The emission core observed at 11.2 and 20.5 $\mu$m shows a 
noticeable North-South 
extension, about 100 pc, while it remains unresolved ($<$ 50 pc) along the 
East-West direction. In the model devised by Granato
et al (1997) of the AGN in NGC\,1068, a 100 pc
torus and a viewing angle of 65\degr\ were the torus parameters finally
selected to match the SED of the AGN. This model predicts indeed 
that maps of the torus 
emission in the mid-IR should be elongated perpendicularly to the 
torus plane, in this case roughly along the North-South direction, and 
on a scale of +/-- $0.5''$. One should notice however that the model 
maps have been obtained after convolution with a PSF FWHM $= 0.09''$,
too narrow compared with the effective PSF of the data 
presented in this paper. Yet, the predicted direction of the elongation 
is consistent with the
observed one at 11.2 and 20.5 $\mu$m and its predicted size has the 
right order of magnitude, if compared to the measured one.\\
Comparing the core at 11.2 and 20.5 $\mu$m to high resolution 
maps at 4.8, 3.5 and 2.2 $\mu$m (Marco \& Alloin 2000, Rouan et al 1998) 
is more difficult because of the 
difference in spatial resolution. Indeed, the extensions 
(both polar and equatorial) seen at 4.8, 3.5 and 2.2 $\mu$m are 
entirely enclosed within the 11.2 and 20.5 $\mu$m core size. 
Yet, if the PA $=$ 102 \degr\ equatorial extension detected at 
4.8, 3.5 and 2.2 $\mu$m features the torus itself, 
one might expect to see emission in the mid-IR from cool dust located 
further away than the warm dust emitting at 4.8 $\mu$m, still along
(PA $=$ 102\degr). Such emission is not detected. One possible 
interpretation is that the equatorial extension at 4.8, 3.5 and 2.2 $\mu$m
does not directly feature the torus. This indeed would be surprising, 
because the covering factor of the equatorial extension to the 
central source is smaller than expected on the basis of several 
arguments (opening angle of the ionizing cone, ratio between IR and 
primary UV radiation...). 
The extended equatorial emission at 4.8, 3.5 and 2.2 $\mu$m could 
instead trace only the equatorial plane of the torus rather than the 
torus itself (possibly smaller) and outline the
merging of the torus with the host--galaxy disc, in regions of high density 
where star formation is occurring and provides an additional and local 
source of heating. This would explain 
the presence of dust at a temperature much higher than that predicted 
by torus models which take only into account the heating from the 
central engine. The dust emission in these regions located in the
torus equatorial plane but "around-the-torus"  
could then be more prominent at 4.8 than at 11.2 and 20.5 $\mu$m. In fact 
there is a whole new area to be explored at the transition between the 
molecular/dusty torus and its environment.

\subsection{Origin of the extended structure}

The comparison of maps in the radio, mid-IR, near-IR and optical/UV
can potentially bring clues about the origin of the mid-IR extended 
emission beyond the torus itself: extra dust components, location, source 
of heating, etc. Unfortunately, one cannot yet confront in detail the 
observed maps 
with predicted model-maps for the
following reason: existing AGN models are targeted at representing the 
molecular/dusty torus itself rather than the full molecular/dusty 
environment of an AGN and therefore do not take into account the presence
of the NLR region or more generally the distribution of matter
away from the torus itself. For example, thermal emission 
from warm dust possibly surviving on the back and UV--protected side 
of NLR clouds is not considered, the impact of the radio jets 
on intervening material such as the NLR clouds or massive molecular
clouds away from the equatorial plane of the torus are not taken into 
account either. 
 
On the contrary, current results on the extended mid-IR emission in 
NGC\,1068 are telling us that material is present and heated around 
and away from the torus itself, up to a distance of $\approx$ 300 pc.
Assuming silicate grains with a power--law (-3.5 exponent) size 
distribution over the size range 0.01 to 0.1 $\mu$m, the 11.2 to 20.5 
$\mu$m flux
ratio observed in cloud\,(a) and cloud\,(c) implies
temperatures of around
150 K. Cloud\,(a) is already more than 200 pc away from the central engine
and on the edge of the ionizing cone: how are the grains heated up 
to this temperature? What is the role played by 
single photon transient heating of small grains? by the radio
jet-like structure which appears to coincide so closely with the mid-IR 
extended
emission? These questions deserve a more complete analysis and a
quantitative modeling which is deferred to a specific and later work.
 
\subsection{Concluding remarks}

Mid-IR imaging at high angular resolution offers potential
advantages in the study of AGN environment because this wavelength range
is specific of warm/cool dust emission (and possibly synchrotron emission 
from electrons) and because extinction is reduced. The
diffraction-limited images (resolution ($0.6''$)) presented in this work
highlight the presence of a prominent 
core emitting about 95\% of the total flux in the mid-IR, as 
well as of extended 
emission, up to $4''$ to the South-West (PA = 210\degr) and $4''$ to the 
North-East 
(PA = 35\degr), broken into patchy components which are particularly 
conspicuous at 20.5 $\mu$m and can be singled-out as
individual clouds. The central core shows an unresolved East-West FWHM 
of $0.6''$ and a North-South FWHM of $0.9''$
corresponding to a resolved full size extension of $\approx$ 100 pc. 
The North-South elongation of the emission core agrees with predicted 
maps of the mid-IR emission from a 100 pc
dusty/molecular torus
surrounding the central engine in NGC\,1068 and observed under an inclination 
angle
of around 65\degr. As a result of smaller optical depth, the extended 
emission in the North-East and 
South-West quadrants is more prominent at
20.5 than at 11.2 $\mu$m. The extended emission  
follows roughly the direction of the radio-jet and radio-lobe structures. 
In the North-East quadrant, the mid-IR emission is located at the eastern 
edge of the ionizing cone outlined by the HST [OIII]--emitting clouds. 
Interpreting the complete molecular/dusty environment of the AGN,
both in the torus and away from it, pleas for the development 
of three-dimensional complex modeling. High resolution imaging 
is the first step in disentangling 
the various components: the new generation of 8m--10m class telescopes 
provides a resolution of $0.3''$ in the mid-IR. Subsequent integral-field 
spectroscopy with such a spatial resolution and interferometry will also
constitute invaluable tools to resolve this type of problem.

\acknowledgements 
We are gratefully indebted to R.Jouan and M.Lortholary for their efficient 
assistance with
the CAMIRAS instrument, as well as to the staff at both ESO/La Silla and 
CFHT/Hawaii for their support during the observing runs. We acknowledge 
careful comments from the referee R.Antonucci.
 
 \end{document}